\documentclass[conference]{IEEEtran}
\usepackage{booktabs}
\usepackage{makecell}
\usepackage{balance}
\usepackage{float}
\usepackage{amsmath,amssymb}
\usepackage{cite}
\usepackage{amsmath,amssymb,amsfonts}
\usepackage{algorithmic}
\usepackage{graphicx}
\usepackage{textcomp}
\usepackage{xcolor}
\usepackage{listings}
\usepackage{algorithm}
\usepackage{algorithmic}
\usepackage{graphicx}
\usepackage{multirow}
\usepackage{amssymb}
\usepackage{tabularx}
\usepackage{booktabs}
\usepackage{array}
\usepackage[utf8]{inputenc}
\usepackage{cite}        
\usepackage{url}  
\usepackage{paralist}
\usepackage{listings}
\usepackage{xcolor}

\lstdefinestyle{mystyle}{
    language=bash,
    basicstyle=\ttfamily\small,
    keywordstyle=\color{blue!70!black}\bfseries,
    commentstyle=\color{green!50!black}\itshape,
    stringstyle=\color{orange!70!black},
    numberstyle=\tiny\color{gray},
    frame=single,
    framesep=1pt,
    rulecolor=\color{black!30},
    backgroundcolor=\color{gray!5},
    tabsize=1,
    captionpos=b,
    showstringspaces=false
}
\usepackage[normalem]{ulem}
\usepackage{tabularx}
\newtheorem{mydef}{Definition}

\def\BibTeX{{\rm B\kern-.05em{\sc i\kern-.025em b}\kern-.08em
    T\kern-.1667em\lower.7ex\hbox{E}\kern-.125emX}}
\begin{document}


\title{ANTMAN: Automatic RTL-Level Feature Extraction and Run-Time Detection of Stealthy Branch Predictor Attacks in the BOOM RISC-V Processor}

\author{
Muhammad Hassan\IEEEauthorrefmark{1},
Sharjeel Imtiaz\IEEEauthorrefmark{1},
Maria Mushtaq\IEEEauthorrefmark{2},\\
Jaan Raik\IEEEauthorrefmark{1}, Tara Ghasempouri\IEEEauthorrefmark{1}
\\
\IEEEauthorrefmark{1}Department of Computer Systems, Tallinn University of Technology, Tallinn, Estonia \\
\IEEEauthorrefmark{2}Télécom Paris, Institut Polytechnique de Paris, Palaiseau, France
\\
\{muhammad.hassan, sharjeel.imtiaz, jaan.raik, tara.ghasempouri\}@taltech.ee\\
\ maria.mushtaq@telecom-paris.fr
}





%
\title{ANTMAN: An Efficient and Interpretable RTL-Level Run-Time Detection Framework for Stealthy Branch Predictor Attacks on BOOM}
\maketitle

\begin{abstract}
Runtime detection of microarchitectural side-channel attacks remains significantly underexplored in RISC-V compared with x86 and ARM ISAs, posing a serious threat to critical applications. State-of-the-art branch predictor attacks bypass traditional data and instruction caches by directly exploiting the state of internal history tables, making them inherently stealthy. Recent research has explored offline detection of microarchitectural attacks on RISC-V; however, efficient runtime detection of microarchitectural attacks on RISC-V hardware remains significantly unaddressed. State-of-the-art hardware-based runtime detection solutions leverage hardware performance counters (HPCs) but suffer from a restricted set of counter registers and tradeoffs between detection accuracy, detection speed, and sampling granularity, making them impractical for stealthy attacks. Moreover, sampling HPCs after distinct intervals leaves intermediate relationships between different microarchitectural blocks unobserved. Additionally, proprietary x86 and ARM ISAs constrain researchers from modifying processor microarchitectural designs. To address these limitations, we propose the first secure-by-design, highly interpretable, non-intrusive, RTL-level runtime detection solution for stealthy branch predictor attacks on BOOM RISC-V, evaluated under both simplified Next-Line Predictor (NLP) and complex TAGE predictor configurations. The attack detection relies on association rules extracted offline and embedded in hardware as a non-intrusive rule monitor that enables runtime detection. The proposed approach achieves excellent detection speed, terminates execution before secret disclosure, and produces zero false positives while remaining flexible for detecting previously unseen variants within the same family of branch predictor attacks.

\end{abstract}


\section{Introduction and Related Work}
Microarchitectural side-channel attacks have gained considerable research attention following the disclosure of the Spectre and Meltdown vulnerabilities. These vulnerabilities exposed hardware design flaws in speculative execution and out-of-order execution that can be exploited through software, posing serious security concerns for security-critical systems\cite{spectre,meltdown}. In the literature, various detection and mitigation solutions have been proposed across different abstraction layers, ranging from operating system (OS)-level approaches\cite{OS} to hardware-based solutions\cite{MAM_MARIA1,MAM_MARIA2}.\\The state-of-the-art hardware-based runtime detection solutions leverage built-in Hardware Performance Counters (HPCs)\cite{MAM_MARIA1, MAM_MARIA2, Carna, ALAM, poly, Choudhari }. Although these initial research efforts to detect Spectre and Meltdown by leveraging HPCs were a pragmatic solution at the time, subsequent research has revealed various limitations, indicating that HPC-based detection still has a long way to go before being considered a practical hardware-based solution\cite{hpc_limitations}.\\HPC-based solutions fundamentally suffer because access is limited to a restricted set of hardware registers, while multiplexing issues further restrict the number of hardware registers that can be sampled simultaneously. Moreover, the coarse sampling granularity compromises detection speed, resulting in a speed-performance trade-off\cite{hpc_limitations, challenges}.\\Moreover, when attacks are hidden within benign applications (camouflaged setup), HPC-based detection solutions cannot detect them\cite{yuval}. These limitations stems from the proprietary and confidential nature of commercial x86 processor microarchitectures, such as Intel and AMD. Consequently, access is limited to a restricted set of HPCs, which provide only aggregated event counts and cannot capture instantaneous processor behavior or internal RTL-level pipeline events.\\ In recent years, RISC-V processors have gained widespread adoption due to their modularity, scalability, and open-source design. Nevertheless, their susceptibility to microarchitectural attacks due to the lack of hardware countermeasures remains a key security concern\cite{RISCV}.
Recent research efforts have presented offline detection solutions using the gem5 computer architecture simulator to detect microarchitectural attacks~\cite{mah1,mah3, palumbo,hassan}. However, runtime detection on real RISC-V hardware remains unexplored. An initial effort to detect microarchitectural attacks on a RISC-V BOOM core leveraged Hardware Performance Counters (HPCs)~\cite{RISCV_HPCs}; however, it inherits the limitations of HPC-based detection.\\
Recently, researchers demonstrated branch predictor attacks on the open-source RISC-V BOOM processor, targeting resource-constrained devices and validating them on an FPGA\cite{MAIN_ATTACK}.
Interestingly, different variants of these branch predictor attacks exhibit  timing differences of less than 20 CPU cycles. 
From a defensive perspective, timing differences of only 5--20 CPU cycles render these attacks inherently stealthy, making them completely indistinguishable from benign workloads using HPC-based detection.\\To address these limitations, we propose the first secure-by-design, non-intrusive, RTL-level runtime detection solution for stealthy attacks on the RISC-V BOOM processor without speed-performance trade-offs, multiplexing issues, or performance overhead. This work addresses the inherent limitations of HPC-based detection by introducing a new secure-by-design research direction. Although the detection of microarchitectural attacks hidden inside benign workloads (camouflaged setup) remains beyond the scope of this paper, the proposed approach provides a strong foundation for future research in this direction.
\vspace{-0.3em}
\subsection{Contributions}
The key contributions of the paper are as follows.
\begin{enumerate}
\item To the best of our knowledge, this paper presents the first secure-by-design, non-intrusive, RTL-level runtime detection method for stealthy microarchitectural attacks on the RISC-V BOOM processor.
\item The proposed detection method is highly interpretable and automatically discovers frequent patterns using association rule mining, identifying the most relevant RTL signals for branch predictor attack detection.

\item An RTL-level runtime detection method that enables the finest possible access to every internal signal of the entire processor pipeline. In addition, the proposed non-intrusive rule-based monitor eliminates the inherent limitations of HPC-based detection, including event multiplexing, limited hardware counter availability, and coarse-grained event visibility, with negligible performance overhead.

\item Finally, we evaluate the proposed detection method across a sequence of computationally intensive, memory-intensive, and branch-heavy workloads, achieving zero false positives while demonstrating its adaptability to a previously unseen branch predictor attack variant.

\end{enumerate}
The remainder of this paper is organized as follows. Section II introduces the key terminology, Section III describes the threat model, Section IV presents the attack model, Section V outlines the proposed methodology, Section VI describes the experimental results, and Section VII concludes the paper and discusses future work.

\section{Preliminaries}
This section introduces the key terminologies used throughout the paper and reviews the related work.

\begin{mydef} \label{DEF1:ARM} \emph{\textbf{Association Rule Mining (ARM)}} discovers association rules from frequent itemsets. An association rule is represented as $X \rightarrow Y$, where $X, Y \subseteq W$ and $X \cap Y = \emptyset$. Here, $W$ denotes the universal set of monitored RTL signals from RISC-V BOOM \cite{rule1,reza}.
\end{mydef}

\begin{mydef} \label{DEF9:FPGrowth}
\emph{\textbf{Frequent Pattern Growth (FP-Growth)}} is one of the most efficient data mining algorithms for finding frequent itemsets from very large transactional datasets. It first compresses the transaction database into an FP-tree (Frequent Pattern Tree) and then extracts frequent patterns directly from the FP-tree.
\end{mydef}

\begin{mydef}
\label{DEF2:Support}
\emph{\textbf{Support}} measures the frequency of occurrence of a frequent itemset and is defined as $\mathrm{Supp}(X)=P(X)$, where $P(X)$ denotes the probability of occurrence of itemset $X$. In this paper, the minimum support threshold used as the input parameter to FP-Growth is set to $\theta_s=5\%$~\cite{support}.
\end{mydef}

\begin{mydef}
\label{DEF3:Confidence}
\emph{\textbf{Confidence}} measures the reliability of an association rule and is defined as $\mathrm{Conf}(X \rightarrow Y)=P(Y \mid X)$. In this paper, the minimum confidence threshold for association rule generation is set to $\theta_c=90\%$~\cite{reza}.
\end{mydef}

\begin{mydef}
\label{DEF4:Chipyard}
\emph{\textbf{Chipyard}} is an open-source RISC-V SoC generator framework integrating configurable RISC-V processor cores, memory subsystems, and peripheral devices.
\end{mydef}

\begin{mydef}
\label{DEF5:Verilator}
\emph{\textbf{Verilator}} is an open-source Verilog/SystemVerilog simulator that compiles RTL designs into optimized C++ models for cycle-accurate simulation~\cite{verilator}.
\end{mydef}

\begin{mydef}
\label{DEF6:BOOM}
\emph{\textbf{BOOM}} is an open-source out-of-order (O3) RISC-V processor core integrated within the Chipyard SoC generator framework~\cite{boom}.
\end{mydef}

\begin{mydef}
\label{DEF7:Bimodal_predictr}
\emph{\textbf{Bimodal Predictor}} is a simple branch predictor that predicts the direction of conditional branches based on their previous outcomes~\cite{boom}.
\end{mydef}

\begin{mydef}
\label{DEF:TAGE_predictor}
\emph{\textbf{Tagged Geometric History Length (TAGE) Predictor}} is a branch predictor that predicts branch directions using multiple prediction tables with different branch history lengths ~\cite{boom}.
\end{mydef}

\section{Threat Model}
Our threat model considers an unprivileged attacker executing malicious code co-located with the victim on the same physical core of an out-of-order (O3) BOOM processor supporting speculative execution. The attacker can manipulate the shared branch predictor to launch branch predictor attacks and leak secret information through side channels but has no direct access to the internal branch predictor state and relies only on observable microarchitectural side effects. We assume a controlled and deterministic bare-metal execution environment and trusted hardware free from physical tampering and hardware Trojans.
\begin{figure*} 
    \centering
    \includegraphics[width=\textwidth]{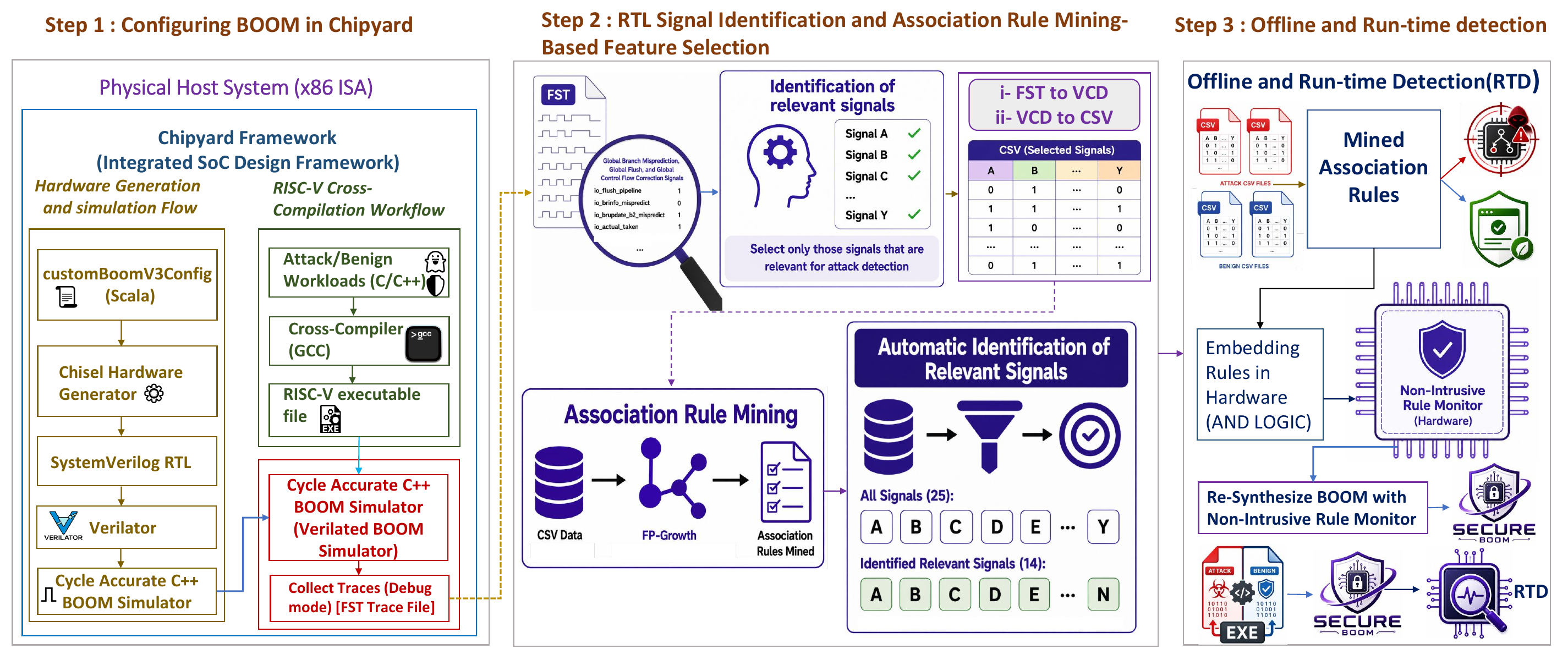}
    \caption{ANTMAN (Secure by design RTL level run-time detection framework)}
    \label{fig:complete_workflow}
\end{figure*} 
\section{Attack Model}

The proposed run-time detection framework is designed for the branch predictor attack model adopted from~\cite{MAIN_ATTACK}, which consists of the following phases:\\
1. \textbf{Training:} The attacker trains the shared branch predictor through repeated branch executions.
   \\
2. \textbf{Victim Execution:} The victim executes secret-dependent branches, updating the shared branch predictor state.
   \\
3. \textbf{Probe:} The attacker probes the predictor state by measuring timing differences to infer secret information.

\section{Methodology}
\label{sec:meth}
In this paper, we propose a comprehensive branch predictor attack detection framework that automatically identifies the most relevant RTL signals using association rule mining and enables offline and run-time detection of stealthy branch predictor attacks on the RISC-V BOOM processor. \\
In \textbf{Step 1} of the proposed framework, the BOOM processor is configured within the Chipyard environment. The customized BOOM configuration is translated into synthesizable SystemVerilog and compiled by Verilator into a cycle-accurate C++ simulation model. The attack and benign workloads are cross-compiled into RISC-V executables and executed on the generated C++ model. Finally, the processor execution traces generated during each workload execution are captured as FST trace files in debug mode. \\
In \textbf{Step 2}, the complete processor behavior during attack execution is available as FST traces. Initially, a set of RTL signals that may be affected by branch predictor attacks is selected based on domain knowledge. The selected FST traces are then converted into CSV format, where association rule mining generates association rules, enabling the automatic identification of the most relevant RTL signals affected during the attack execution.\\
Finally, in \textbf{Step 3}, the association rules generated in Step 2 are used for offline attack detection across computational, memory-intensive, and branch-heavy workloads. The generated rules are then implemented in hardware as a non-intrusive rule monitor, where the detector triggers an assertion upon rule matching to detect stealthy branch predictor attacks at run time. The detailed implementation of each step illustrated in Fig.~\ref{fig:complete_workflow} is as follows.\\
\textbf{Step~1: Configuring BOOM in Chipyard}
\\
The first step of the proposed ANTMAN framework is to configure the BOOM processor within the Chipyard environment. Chipyard is a RISC-V-based system-on-chip (SoC) generation framework that provides an integrated environment for hardware generation, simulation, and software cross-compilation using the RISC-V GNU toolchain. In Chipyard, BOOM configuration and workload cross-compilation are performed as two independent phases. During the BOOM configuration phase, the same BOOM parameters used in \cite{mah_paper} are adopted, as listed in row~2 of Table~\ref{tab:Exp-Setup}.
\begin{table}[t]
\centering
\caption{Experimental Setup and Configuration}
\label{tab:Exp-Setup}
\begin{tabular}{p{0.25\linewidth} p{0.650\linewidth}}
\hline
\textbf{Item} & \textbf{Description} \\
\hline

\textbf{Chipyard-Based Experimental Platform} &
Chipyard SoC generator framework\newline
BOOM v3 (Scaled Configuration)\newline
RV64GC ISA\newline
Verilator (Compiles Verilog/SystemVerilog RTL into cycle-accurate C++ models)\newline
RISC-V GNU Cross-Compilation Toolchain (\texttt{riscv64-unknown-elf-gcc}) \\
\hline

\textbf{BOOM \hspace{4em} Configuration~\cite{mah_paper}} &
Reorder Buffer: 48 entries\newline
Integer IQ: 16 entries\newline
Memory IQ: 10 entries\newline
FP IQ: 12 entries\newline
Integer PRF: 64 registers\newline
FP PRF: 48 registers\newline
Load Queue: 12 entries\newline
Store Queue: 12 entries\newline
Fetch Target Queue: 32 entries\newline
L1D Cache: 16 KB, 4-way, 64-B line\newline
L1I Cache: 16 KB, 4-way, 64-B line \\
\hline

\textbf{Benign Workloads} &
dhrystone, mm, median, qsort, spmv, towers, vvadd \\
\hline

\textbf{Attack Workloads} &
CBPA, IBPA \\
\hline

\textbf{Timing \hspace{4em} Configuration} &
500 attack rounds using \texttt{rdcycle}-based latency measurements. \\
\hline

\end{tabular}
\end{table}
These BOOM configurations are first translated into a synthesizable SystemVerilog model of BOOM. Subsequently, Verilator translates this model into a cycle-accurate C++ model of BOOM. In the workload preparation phase, the inputs to this C++ model are generated by cross-compiling the source codes of the branch predictor attack workloads and a set of benign RISC-V benchmarks. The resulting cross-compiled executable files are provided as inputs to the BOOM model. Execution of these workload files in debug mode captures the complete end-to-end behavior of the CPU pipeline for each workload in FST waveform format. \\
Table~\ref{tab:Exp-Setup} summarizes the complete experimental setup and BOOM configurations.\\
\textbf{Step~2: RTL Signal Identification and Association Rule Mining-Based Feature Selection}\\
The output of Step~1 captures the complete CPU pipeline execution behavior under the attack workload in the form of an FST trace file. This step consists of two key phases: (a) RTL signal identification and (b) association rule mining-based feature selection. The following discussion briefly summarizes each phase.
\subsection{RTL Signal Identification}
In the first phase of Step~1, a set of RTL signals relevant to branch predictor attacks are selected from different microarchitectural blocks of the BOOM processor based on the behavior of branch predictor attacks. As listed in Table~\ref{tab:rtl_signals}, these signals belong to microarchitectural blocks involved in branch prediction and recovery.\\ The Frontend Branch Prediction Block, which combines the Next Line Predictor (NLP) and Branch Target Buffer (BTB), predicts the next instruction fetch address and provides branch target information during speculative instruction fetch. The BPD (Backing Predictor Direction) block predicts the branch direction (Taken/Not Taken) using predictors such as Bimodal or TAGE. Similarly, the Global History Register (GHR) block maintains and tracks previous branch outcomes and provides branch history information to improve future direction predictions. \\ The Branch Resolution/Execution block determines the actual branch outcome during execution and compares it with the earlier speculative prediction to detect mispredictions. When a branch misprediction occurs, this information is propagated across microarchitectural blocks to initiate control-flow recovery. The Branch Update Broadcast block distributes branch resolution outcomes and misprediction information to the frontend and other microarchitectural blocks to coordinate recovery actions. The Frontend Redirect Control block redirects instruction fetch to the correct Program Counter (PC), while
\begin{table}
\caption{Categorization of Selected RTL Signals}
\label{tab:rtl_signals}
\centering
\renewcommand{\arraystretch}{1.15}
\begin{tabular}{p{3.2cm} p{5.0cm}}
\hline
\textbf{Microarchitectural Block} & 
\textbf{Associated RTL Signals} \\ 
\hline

Global History Register (GHR) &
\texttt{ghist\_s1\_branch\_not\_taken},
\texttt{ghist\_s2\_branch\_not\_taken},
\texttt{ifu\_redirect\_ghist\_saw\_nt}
\\

NLP / BTB (Next Line Predictor / Branch Target Buffer) &
\texttt{f3\_btb\_misp0},
\texttt{f3\_btb\_misp1},
\texttt{f3\_btb\_misp2},
\texttt{f3\_btb\_misp3}
\\

BPD (Backing Predictor Direction) &
\texttt{bpd\_is\_mispredict\_update}
\\

Branch Resolution / Execute &
\texttt{cpu\_brupdate\_b2\_misp},
\texttt{exu\_brinfo\_misp},
\texttt{brinfo0\_misp},
\texttt{b2\_misp}
\\

Frontend Redirect Control &
\texttt{redirect\_val},
\texttt{redirect\_flush}
\\

Branch Update Broadcast Network &
\texttt{ifu\_brupdate\_b2\_misp},
\texttt{lsu\_brupdate\_b2\_misp},
\texttt{fp\_rename\_brupdate\_b2\_misp},
\texttt{rob\_brupdate\_b2\_misp}
\\

Pipeline Flush / Recovery Control &
\texttt{rob\_flush\_valid},
\texttt{rob\_flush\_frontend},
\texttt{dec\_flush\_pipeline},
\texttt{mem\_issue\_flush\_pipeline},
\texttt{int\_issue\_flush\_pipeline},
\texttt{fp\_pipe\_flush\_pipeline},
\texttt{alu\_queue\_flush}
\\

\hline
\end{tabular}
\end{table}
the Pipeline Flush/Recovery Control block flushes incorrect-path instructions and their dependent operations from queues and pipeline stages, ultimately restoring the processor state for correct execution. \\The FST trace file containing the selected RTL signals is first converted into VCD format and subsequently exported to CSV format to facilitate further analysis and extract meaningful insights. Table~\ref{tab:rtl_signals} summarizes the selected RTL signals.

\subsection{Association Rule Mining-Based Feature Selection}
In Phase~2 of Step~2, after identifying the relevant RTL signals and capturing their complete execution behavior in a CSV file, the next important step is to apply association rule mining to discover relationships and patterns among the selected signals. This stage plays a crucial role, as humans cannot directly identify complex patterns among multiple signals across millions of clock cycles, considering signal variations and their interdependencies. 
Therefore, we opted for Frequent Pattern Growth (FP-Growth), defined in Definition~\ref{DEF9:FPGrowth}, to efficiently identify hidden patterns in the execution of branch-predictor attacks. From the data mining output, we obtained a set of 16 interpretable association rules based on predefined minimum support \textit{5\%} and confidence \textit{90\%} thresholds, as defined in Definition~\ref{DEF2:Support} and Definition~\ref{DEF3:Confidence}, respectively.
\\
From these association rules, we observe a highly structured and repeated coupling behavior among misprediction, redirect (control-flow correction), and full-pipeline flush signals. Three dominant behavioral clusters are identified: Global Branch Misprediction, Global Redirect (Control-Flow Correction), and Global Pipeline Flush. The following discussion presents a detailed analysis of each cluster.
\subsubsection{Global Branch Misprediction}
Table~\ref{tab:global_branch_misprediction} presents the first cluster, which reveals strong coupling among branch 

\setlength{\tabcolsep}{3pt}

\begin{table*}[!t]
\caption{Global Branch Misprediction Association Rules}
\label{tab:global_branch_misprediction}
\centering
\footnotesize

\begin{tabularx}{0.90\textwidth}
{
>{\raggedright\arraybackslash}p{0.079\textwidth}
>{\raggedright\arraybackslash}X
>{\raggedright\arraybackslash}p{0.21\textwidth}
}
\toprule
\textbf{Rule} & \textbf{Antecedent} & \textbf{Consequent} \\
\midrule

Rule 1 &
cpu\_brupdate\_b2\_misp AND ifu\_brupdate\_b2\_misp AND b2\_misp AND fp\_rename\_brupdate\_b2\_misp AND lsu\_brupdate\_b2\_misp AND redirect\_val AND redirect\_flush
&
rob\_brupdate\_b2\_misp
\\

Rule 2 &
cpu\_brupdate\_b2\_misp AND rob\_brupdate\_b2\_misp AND b2\_misp AND fp\_rename\_brupdate\_b2\_misp AND lsu\_brupdate\_b2\_misp AND redirect\_val AND redirect\_flush
&
ifu\_brupdate\_b2\_misp
\\

Rule 3 &
rob\_brupdate\_b2\_misp AND ifu\_brupdate\_b2\_misp AND b2\_misp AND fp\_rename\_brupdate\_b2\_misp AND lsu\_brupdate\_b2\_misp AND redirect\_val AND redirect\_flush
&
cpu\_brupdate\_b2\_misp
\\

Rule 4 &
cpu\_brupdate\_b2\_misp AND rob\_brupdate\_b2\_misp AND ifu\_brupdate\_b2\_misp AND b2\_misp AND fp\_rename\_brupdate\_b2\_misp AND redirect\_val AND redirect\_flush
&
lsu\_brupdate\_b2\_misp
\\

Rule 5 &
cpu\_brupdate\_b2\_misp AND rob\_brupdate\_b2\_misp AND ifu\_brupdate\_b2\_misp AND fp\_rename\_brupdate\_b2\_misp AND lsu\_brupdate\_b2\_misp AND redirect\_val AND redirect\_flush
&
b2\_misp
\\

Rule 6 &
cpu\_brupdate\_b2\_misp AND rob\_brupdate\_b2\_misp AND ifu\_brupdate\_b2\_misp AND b2\_misp AND lsu\_brupdate\_b2\_misp AND redirect\_val AND redirect\_flush
&
fp\_rename\_brupdate\_b2\_misp
\\

\bottomrule
\end{tabularx}
\end{table*}

\begin{table*}[!t]
\caption{Global Control Flow Correction Association Rules}
\label{tab:global_control_flow}
\centering
\footnotesize

\begin{tabularx}{0.90\textwidth}
{
>{\raggedright\arraybackslash}p{0.079\textwidth}
>{\raggedright\arraybackslash}X
>{\raggedright\arraybackslash}p{0.21\textwidth}
}
\toprule
\textbf{Rule} & \textbf{Antecedent} & \textbf{Consequent} \\
\midrule

Rule 1 &
dec\_flush\_pipeline AND int\_issue\_flush\_pipeline AND alu\_queue\_flush AND mem\_issue\_flush\_pipeline AND fp\_pipe\_flush\_pipeline AND ifu\_redirect\_ghist\_saw\_nt AND redirect\_flush
&
redirect\_val
\\

Rule 2 &
cpu\_brupdate\_b2\_misp AND rob\_brupdate\_b2\_misp AND ifu\_brupdate\_b2\_misp AND b2\_misp AND fp\_rename\_brupdate\_b2\_misp AND lsu\_brupdate\_b2\_misp AND redirect\_flush
&
redirect\_val
\\

Rule 3 &
cpu\_brupdate\_b2\_misp AND rob\_brupdate\_b2\_misp AND ifu\_brupdate\_b2\_misp AND b2\_misp AND fp\_rename\_brupdate\_b2\_misp AND lsu\_brupdate\_b2\_misp AND redirect\_val
&
redirect\_flush
\\

Rule 4 &
dec\_flush\_pipeline AND int\_issue\_flush\_pipeline AND alu\_queue\_flush AND mem\_issue\_flush\_pipeline AND fp\_pipe\_flush\_pipeline AND ifu\_redirect\_ghist\_saw\_nt AND redirect\_val
&
redirect\_flush
\\

\bottomrule
\end{tabularx}
\end{table*}
\begin{table*}[!t]
\caption{Global Recovery Association Rules}
\label{tab:global_recovery}
\centering
\footnotesize

\begin{tabularx}{0.90\textwidth}
{
>{\raggedright\arraybackslash}p{0.079\textwidth}
>{\raggedright\arraybackslash}X
>{\raggedright\arraybackslash}p{0.21\textwidth}
}
\toprule
\textbf{Rule} & \textbf{Antecedent} & \textbf{Consequent} \\
\midrule

Rule 1 &
cpu\_brupdate\_b2\_misp AND rob\_brupdate\_b2\_misp AND ifu\_brupdate\_b2\_misp AND b2\_misp AND lsu\_brupdate\_b2\_misp AND redirect\_val AND redirect\_flush
&
fp\_rename\_brupdate\_b2\_misp
\\

Rule 2 &
int\_issue\_flush\_pipeline AND alu\_queue\_flush AND mem\_issue\_flush\_pipeline AND fp\_pipe\_flush\_pipeline AND ifu\_redirect\_ghist\_saw\_nt AND redirect\_val AND redirect\_flush
&
dec\_flush\_pipeline
\\

Rule 3 &
dec\_flush\_pipeline AND int\_issue\_flush\_pipeline AND mem\_issue\_flush\_pipeline AND fp\_pipe\_flush\_pipeline AND ifu\_redirect\_ghist\_saw\_nt AND redirect\_val AND redirect\_flush
&
alu\_queue\_flush
\\

Rule 4 &
dec\_flush\_pipeline AND alu\_queue\_flush AND mem\_issue\_flush\_pipeline AND fp\_pipe\_flush\_pipeline AND ifu\_redirect\_ghist\_saw\_nt AND redirect\_val AND redirect\_flush
&
int\_issue\_flush\_pipeline
\\

Rule 5 &
dec\_flush\_pipeline AND int\_issue\_flush\_pipeline AND alu\_queue\_flush AND fp\_pipe\_flush\_pipeline AND ifu\_redirect\_ghist\_saw\_nt AND redirect\_val AND redirect\_flush
&
mem\_issue\_flush\_pipeline
\\

Rule 6 &
dec\_flush\_pipeline AND int\_issue\_flush\_pipeline AND alu\_queue\_flush AND mem\_issue\_flush\_pipeline AND ifu\_redirect\_ghist\_saw\_nt AND redirect\_val AND redirect\_flush
&
fp\_pipe\_flush\_pipeline
\\

Rule 7 &
dec\_flush\_pipeline AND int\_issue\_flush\_pipeline AND alu\_queue\_flush AND mem\_issue\_flush\_pipeline AND fp\_pipe\_flush\_pipeline AND redirect\_val AND redirect\_flush
&
ifu\_redirect\_ghist\_saw\_nt
\\
\bottomrule
\end{tabularx}
\end{table*}
misprediction signals across multiple microarchitectural blocks. Multiple units, including the Reorder Buffer (ROB), Instruction Fetch Unit (IFU), Load-Store Unit (LSU), and Floating-Point Rename Unit (FP rename), simultaneously detect misprediction events along with control-flow correction signals (\texttt{redirect\_valid} and \texttt{redirect\_flush}), indicating coordinated pipeline-wide misprediction detection and recovery.

  \begin{figure*} 
    \centering
    \includegraphics[width=\textwidth]{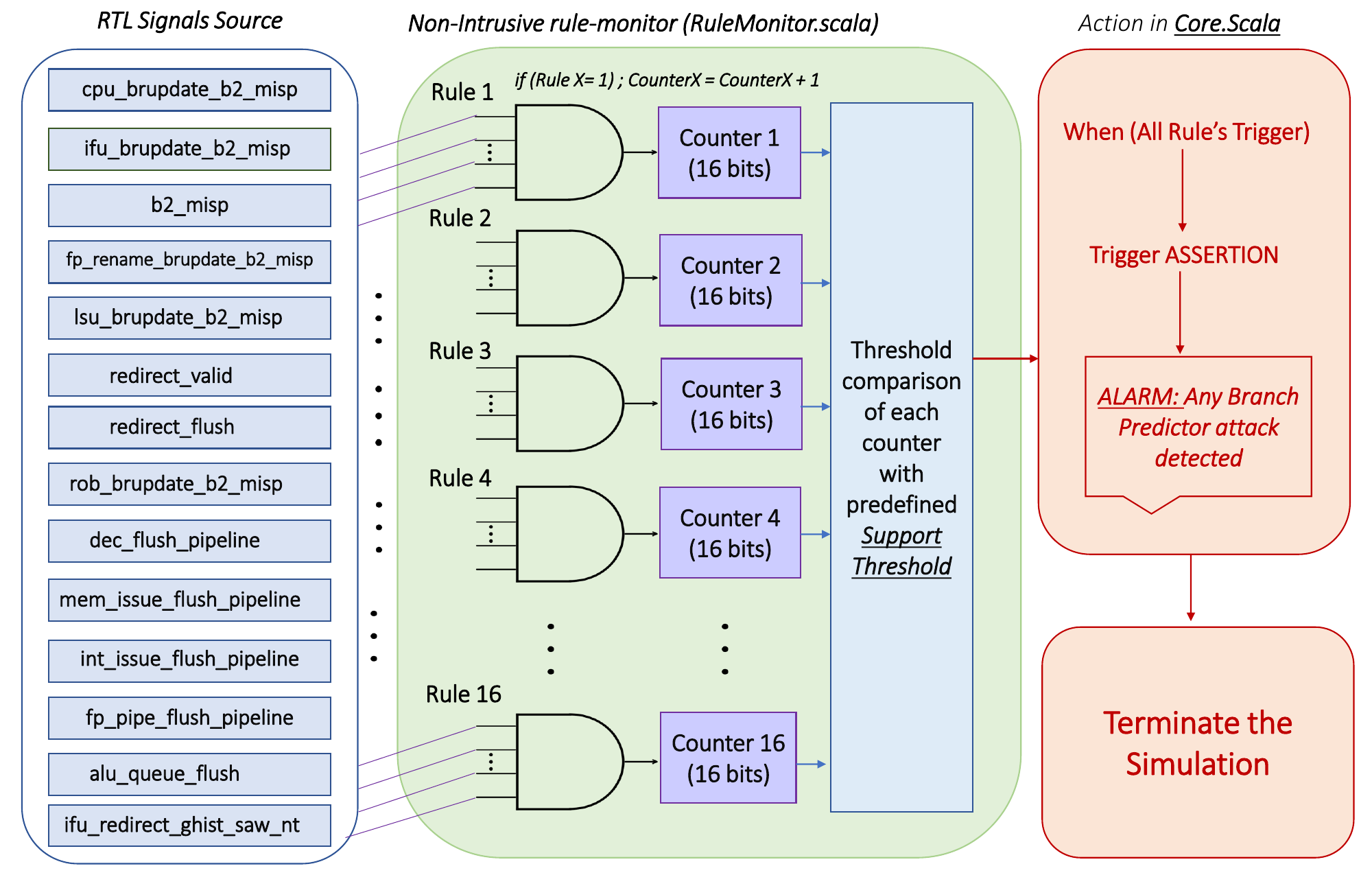}
    \caption{Non-Intrusive Rule Monitor}
    \label{fig:ARM}
\end{figure*} 

\subsubsection{Redirect (Control-flow correction)}
The second cluster, shown in Table~\ref{tab:global_control_flow}, reveals strong coupling between redirect signals and multiple microarchitectural blocks. The mined association rules show that \texttt{redirect\_valid} and \texttt{redirect\_flush} are strongly associated with pipeline flush and branch misprediction events across different \\
microarchitectural blocks, enabling coordinated control-flow recovery. When speculation fails, redirect decisions redirect instruction fetch to the correct path and initiate recovery by propagating control-flow correction across the processor pipeline.

\subsubsection{Global Pipeline Flush (Recovery)}

The association rules in the third cluster (Table~\ref{tab:global_recovery}) highlight strong coordination among pipeline flush signals across multiple microarchitectural blocks. The association rules show that flush events in the decode, issue, execution, memory, and floating-point units are strongly associated with  \texttt{redirect\_valid} and \texttt{redirect\_flush} indicating coordinated pipeline recovery activity following control-flow redirection.
\\
\textbf{Step~3: Offline and Run-time detection}\\
In the final step (Step 3), the extracted rules were first evaluated offline using benign and attack workloads under minimum support and confidence thresholds of 5\% and 90\%, respectively. After validation, the rules were implemented and synthesized as a non-intrusive hardware monitor that tracks rule activations and stops execution when the activation count exceeds the prescribed support threshold, enabling hardware-level detection of stealthy branch predictor attacks. 
From these association rules, we obtain a subset of pre-identified signals across different microarchitectural blocks that are most relevant to branch predictor attack detection. Consequently, the association-rule mining indirectly performs feature selection by identifying the most relevant signals affected during attack execution, thereby enabling efficient run-time detection capability.
\\
Figure 2 shows the integration of the rule-based monitor into the BOOM RISC-V microarchitecture. It can be observed that only 14 automatically discovered RTL signals were tapped from different microarchitectural blocks, and the association rules were converted into hardware using AND logic without causing any interference with ongoing processor operations. The rule-monitoring hardware was implemented in a separate `RuleMonitor.scala' module, which was instantiated in `Core.scala', and the design was subsequently re-synthesized. These embedded rule monitors count every time an association rule is triggered and store the corresponding count for each rule in a dedicated 16-bit counter. If the trigger count of all association rules exceeds the predefined support threshold, an assertion is raised at run time, and program execution is immediately terminated.
\section{Experimental Results}
In this section, we present a systematic evaluation of the proposed detection framework to detect stealthy branch predictor attacks across a sequence of computational, memory-intensive, and branch-heavy workloads. The systematic evaluation first validates the detection framework in an offline environment using only the simplified Next-Line Predictor (NLP) configuration. Once the detection framework is validated using Association Rule Mining (ARM), we carry out run-time detection using both the simplified NLP and the more complex TAGE predictor configuration to evaluate the effectiveness of the proposed detector at run time. The evaluation also considers the effectiveness and practical aspects of the proposed non-intrusive run-time detection framework against stealthy attacks in RISC-V, including detection speed, interpretability, detection accuracy, and performance overhead, and compares the proposed framework with state-of-the-art detection solutions.
\subsection{Offline Detection}
This section aims to validate the effectiveness of Association Rule Mining with the simplified NLP configuration in an offline environment. The association rules are generated using one variant of branch predictor attack (CBPA) and evaluated against both attack variants (CBPA and IBPA) alongside all benign applications. It can be seen from Listing 1 that, with the simplified NLP configuration, both variants of branch predictor attacks can be detected with no false positives. 
\subsection{Run-time detection}
After validating the effectiveness of association rule mining in detecting stealthy branch predictor attacks, we embed the extracted rules into the RISC-V BOOM hardware as a non-intrusive monitor to detect these attacks at run-time. In this section, we consider both variants of branch predictor attacks (CBPA and IBPA) against simplified NLP and TAGE predictor configurations. The benign workloads remain the same for each configuration, consisting of computational, memory-intensive, and branch-heavy workloads used for both predictor configurations. 
\begin{lstlisting}[
    style=mystyle,
    caption={Offline branch predictor attack detection using Association Rule Mining (ARM) with a simplified next-line predictor configuration.} ,
    label={lst:lst1}
]
ANALYZING WORKLOAD TRACES: CBPA.csv
Summary for CBPA.csv
Rules Triggered : 16/16 (100%)

Prediction:
High probability that the branch predictor
is under attack.

-----------------------------------------------
ANALYZING WORKLOAD TRACES: dhrystone.csv

Summary for dhrystone.csv
Rules Triggered : 0/16 (0%)

Prediction:
Very low probability of an attack.

-----------------------------------------------
ANALYZING WORKLOAD TRACES: mm.csv

Summary for mm.csv
Rules Triggered : 0/16 (0%)

Prediction:
Very low probability of an attack.

-----------------------------------------------

ANALYZING WORKLOAD TRACES: qsort.csv

Summary for qsort.csv
Rules Triggered : 0/16 (0%)

Prediction:
Very low probability of an attack.

-----------------------------------------------
ANALYZING WORKLOAD TRACES: median.csv

Summary for median.csv
Rules Triggered : 0/16 (0%)

Prediction:
Very low probability of an attack.

-----------------------------------------------
ANALYZING WORKLOAD TRACES: spmv.csv

Summary for spmv.csv
Rules Triggered : 0/16 (0%)

Prediction:
Very low probability of an attack.


-----------------------------------------------

ANALYZING WORKLOAD TRACES: towers.csv

Summary for towers.csv
Rules Triggered : 0/16 (0%)

Prediction:
Very low probability of an attack.

-----------------------------------------------
ANALYZING WORKLOAD TRACES: vvadd.csv

Summary for vvadd.csv
Rules Triggered : 0/16 (0%)

Prediction:
Very low probability of an attack.
-----------------------------------------------
ANALYZING WORKLOAD TRACES: IBPA.csv
Summary for IBPA.csv
Rules Triggered : 16/16 (100%)

Prediction:
High probability that the branch predictor
is under attack.
\end{lstlisting}

\begin{lstlisting}[
    style=mystyle,
caption={Run-time detection of CBPA and IBPA attacks with the simplified NLP configuration.},
    label={lst:attack_timing},
    captionpos=b
]
[UART] UART0 is here (stdin/stdout).

BRAD-v1 timing test starting...
SECRET_DATA_LEN = 16
ATTACK_ROUNDS   = 500

Index 0 | true=1 | avg(spy0)=97 cycles
        | avg(spy1)=78 cycles
        | faster=spy1 -> guess=1

Index 1 | true=0 | avg(spy0)=76 cycles
        | avg(spy1)=99 cycles
        | faster=spy0 -> guess=0

Index 2 | true=0 | avg(spy0)=76 cycles
        | avg(spy1)=99 cycles
        | faster=spy0 -> guess=0



Index 3 | true=1 | avg(spy0)=97 cycles
        | avg(spy1)=78 cycles
        | faster=spy1 -> guess=1

Assertion failed:
Branch prediction related attack detected.
Simulation terminated.

-----------------------------------------------

[UART] UART0 is here (stdin/stdout).

IBPA with indirect jump helper (rdcycle timing)

Secret:
1 0 0 0 1 0 1 0 1 1 1 0 0 0 1 0

ATTACK_ROUNDS = 500

Index 0 : mean_probe = 97 cycles
Index 1 : mean_probe = 105 cycles
Index 2 : mean_probe = 105 cycles
Index 3 : mean_probe = 97 cycles

Assertion failed:
Branch prediction related attack detected.
Simulation aborted.
\end{lstlisting}
\begin{lstlisting}[
style=mystyle,
caption={Run-time detection of CBPA and IBPA attacks under the complex TAGE configuration},
label={lst:attack_timing_complex},
captionpos=b
]
[UART] UART0 is here (stdin/stdout).

BRAD-v1 timing test starting...
SECRET_DATA_LEN = 16
ATTACK_ROUNDS   = 500

Index 0 | true=1 | avg(spy0)=76 cycles
| avg(spy1)=100 cycles
| faster=spy0 -> guess=0

Index 1 | true=0 | avg(spy0)=76 cycles
| avg(spy1)=100 cycles
| faster=spy0 -> guess=0

Index 2 | true=0 | avg(spy0)=98 cycles
| avg(spy1)=78 cycles
| faster=spy1 -> guess=1

Index 3 | true=1 | avg(spy0)=97 cycles
| avg(spy1)=78 cycles
| faster=spy1 -> guess=1

Assertion failed:
Any branch prediction related attack detected.
Simulation terminated.

-----------------------------------------------

[UART] UART0 is here (stdin/stdout).
IBPA with indirect jump helper (rdcycle timing)
Secret:
1 0 0 1 0 1 0 1 1 1 0 0 1 0 1 0
ATTACK_ROUNDS = 500
Index 0 : mean_probe = 97 cycles
Index 1 : mean_probe = 105 cycles
Index 2 : mean_probe = 105 cycles
Index 3 : mean_probe = 97 cycles
Index 4 : mean_probe = 105 cycles
Index 5 : mean_probe = 97 cycles
Index 6 : mean_probe = 105 cycles
Index 7 : mean_probe = 97 cycles
Index 8 : mean_probe = 97 cycles
Index 9 : mean_probe = 97 cycles

Assertion failed:
Any branch prediction related attack detected.
Simulation aborted.
\end{lstlisting}
From Listings 2 and 3, it can be observed that the ARM-based detector successfully detects and terminates attacks at run-time. Under the simplified NLP configuration, both CBPA and IBPA attacks are detected after 25\% of the attack execution. As reported in \cite{mah_paper}, the TAGE predictor configuration only reduces the probability of successfully recovering the secret. Our simulation results validate this observation, showing an approximately 50\% success rate for secret recovery under the TAGE configuration for CBPA.\\ However, existing cryptanalytic techniques have demonstrated that recovering 50\% of the cryptographic keys through side-channel leakage can enable extraction of the remaining key information. Therefore, the proposed ARM-based detector demonstrates that stealthy attacks under complex TAGE configurations remain detectable at the RTL level.
Under the TAGE configuration, CBPA is detected at 25\% of attack completion with a 50\% secret recovery success rate, whereas the unseen IBPA variant is detected at 56.25\% of completion. Hence, the proposed RTL-level non-intrusive monitor provides an additional hardware security layer, enabling a secure-by-design solution for run-time detection of stealthy attacks across simple and complex branch predictor configurations.
\\
Similarly, for all benign workloads, under both simplified NLP and complex TAGE configurations, no benign workload is terminated during execution, resulting in zero false positives. This detection capability is enabled by association rules that capture structured attack patterns comprising global branch

\setlength{\tabcolsep}{2pt}
\renewcommand{\arraystretch}{1.2}
\begin{table*}[h]
\caption{Comparison of the Proposed Framework (ANTMAN) with State-of-the-Art Approaches}
\label{tab:comparison_framework}
\centering
\footnotesize

\begin{tabularx}{0.90\textwidth}
{
>{\raggedright\arraybackslash}p{0.16\textwidth}
*{9}{>{\centering\arraybackslash}X}
}
\toprule

\textbf{Literature} &
\textbf{\makecell{ISA}} &
\textbf{\makecell{Sampling\\Source}} &
\textbf{\makecell{Secure by \\ Design}} &
\textbf{\makecell{Offline /\\Runtime}} &
\textbf{\makecell{Stealthy\\Attack}} &
\textbf{\makecell{Detection\\Speed}} &
\textbf{Accuracy} &
\textbf{Flexibility} &
\textbf{Interpretability}
\\

\midrule

Mushtaq et al. \cite{MAM_MARIA1}&
x86 &
HPCs &
No &
Runtime &
No &
Yes &
Yes &
No &
Low
\\

Mushtaq et al. \cite{MAM_MARIA2} &
x86 &
HPCs &
No &
Runtime &
No &
Yes &
Yes &
No &
Low
\\

Carnà et al.\cite{Carna} &
x86 &
HPCs &
No &
Runtime &
No &
Yes &
Yes &
No &
Low
\\

Alam et al.\cite{ALAM} &
x86 &
HPCs &
No &
Runtime &
No &
Yes &
Yes &
No &
Low
\\
Polychronou et al. \cite{poly}&
x86 &
HPCs &
No &
Runtime &
No &
Yes &
Yes &
No &
Low
\\Choudhari et al. \cite{Choudhari}&
x86 &
HPCs &
No &
Runtime &
No &
Yes &
Yes &
No &
Low
\\Awais et al. \cite{MAIP}&
x86 &
gem5 &
No &
Offline &
No &
No &
Yes &
No &
Low
\\Khan et al. \cite{MKIP}&
RISC-V &
gem5 &
No &
Offline &
No &
No &
Yes &
No &
Low
\\Palumbo \cite{palumbo}&
RISC-V &
gem5 &
No &
Offline &
No &
No &
Yes &
No &
Low

\\Hassan et al. \cite{hassan}&
RISC-V &
gem5 &
No &
Offline &
No &
No &
Yes &
Yes &
High
\\
Le et al. \cite{Le_et_al}&
RISC-V &
HPCs &
No &
N/A &
No &
No &
Yes &
No &
Low
\\
Proposed Framework (ANTMAN)&
RISC-V &
RTL traces &
Yes &
Runtime &
Yes &
Yes &
Yes &
Yes &
High
\\
\hline
\end{tabularx}
\end{table*}
mispredictions across multiple microarchitectural blocks, control-flow correction, and systematic recovery through flushing dependent instructions across the processor pipeline. Table \ref{tab:attack_matrix} summarizes the attack detection points, expressed as the percentage of attack completion, under different branch predictor configurations.

\begin{table} [htpb]
\caption{Attack Detection points Under Different Branch Predictor Configurations}
\label{tab:attack_matrix}
\centering
\renewcommand{\arraystretch}{1.2}
\begin{tabular*}{0.90\columnwidth}{@{\extracolsep{\fill}}l|l|l}
\hline
\textbf{Attack} & \textbf{Simplified NLP} & \textbf{Complex TAGE} \\
\hline
CBPA & 25\% & 25\% \\
IBPA & 25\% & 56.25\% \\
\hline
\end{tabular*}
\end{table}

\subsection{Comparison with State-of-the-art}
The following table \ref{tab:comparison_framework} provides a systematic comparison of the proposed overall framework with state-of-the-art.

From Table \ref{tab:comparison_framework}, it is evident that the proposed detection framework is the first RTL-level, secure-by-design, run-time attack detection framework enabled by the open-source nature of the RISC-V architecture. The framework is capable of detecting stealthy branch predictor attacks before attack completion, while also providing high interpretability and the flexibility to identify new variants within the same attack family. In contrast, state-of-the-art run-time detection approaches that rely on hardware performance counters (HPCs) suffer from inherent limitations, including restricted visibility into microarchitectural events and dependence on proprietary processor implementations, where only a limited set of events can be monitored through HPC sampling.

\section{Conclusion and Future work}
As open-source RISC-V processors become ubiquitous across diverse applications, including critical systems, their security remains a key concern. However, the open-source nature of RISC-V provides a unique opportunity for the research community to architect efficient, secure-by-design solutions. This work presents the first secure-by-design, interpretable, association rule-based, non-intrusive RTL-level runtime detection method for stealthy branch predictor attacks, validated on BOOM RISC-V. The proposed detection method enables efficient runtime detection without the inherent limitations of HPC-based approaches. Evaluation under simplified Next-Line Predictor (NLP) and complex TAGE predictor configurations demonstrates fast detection, termination before secret disclosure, and flexibility in detecting previously unseen variants within the same branch predictor attack family. The robustness of the approach is further validated across computational, memory-intensive, and branch-heavy workloads, achieving zero false positives while ensuring runtime detection.\\
From a future research perspective, the proposed framework establishes a foundation for extending this end-to-end proof-of-concept to a broader range of microarchitectural attacks on RISC-V and enabling efficient mitigation solutions. A windowed approach can further be introduced to track association-rule activations during stochastic workload execution. This approach offers two key advantages: tracking the number of rule activations and passively monitoring execution duration across windows to identify patterns that exceed predefined support thresholds, thereby enabling non-intrusive detection of attacks hidden within benign workloads. Finally, a detailed power, performance, and area (PPA) analysis can quantify resource utilization and evaluate performance and area overheads prior to fabrication.
\section*{Acknowledgment}
\begin{itemize}
    \item This work was supported by the Estonian Research Council grants PSG837.
    \item The icons and graphical elements used in this work were generated with assistance from OpenAI's ChatGPT image generation tool. The generated images were subsequently reviewed, modified, and adapted by the authors for inclusion in this manuscript.
\end{itemize}

\newpage

\bibliography{ANTMAN/ANTMAN}         

\begin{thebibliography}{10}
\providecommand{\url}[1]{#1}
\csname url@samestyle\endcsname
\providecommand{\newblock}{\relax}
\providecommand{\bibinfo}[2]{#2}
\providecommand{\BIBentrySTDinterwordspacing}{\spaceskip=0pt\relax}
\providecommand{\BIBentryALTinterwordstretchfactor}{4}
\providecommand{\BIBentryALTinterwordspacing}{\spaceskip=\fontdimen2\font plus
\BIBentryALTinterwordstretchfactor\fontdimen3\font minus \fontdimen4\font\relax}
\providecommand{\BIBforeignlanguage}[2]{{%
\expandafter\ifx\csname l@#1\endcsname\relax
\typeout{** WARNING: IEEEtran.bst: No hyphenation pattern has been}%
\typeout{** loaded for the language `#1'. Using the pattern for}%
\typeout{** the default language instead.}%
\else
\language=\csname l@#1\endcsname
\fi
#2}}
\providecommand{\BIBdecl}{\relax}
\BIBdecl

\bibitem{spectre}
\BIBentryALTinterwordspacing
P.~Kocher, J.~Horn, A.~Fogh, D.~Genkin, D.~Gruss, W.~Haas, M.~Hamburg, M.~Lipp, S.~Mangard, T.~Prescher, M.~Schwarz, and Y.~Yarom, ``Spectre attacks: exploiting speculative execution,'' \emph{Commun. ACM}, vol.~63, no.~7, p. 93–101, Jun. 2020. [Online]. Available: \url{https://doi.org/10.1145/3399742}
\BIBentrySTDinterwordspacing

\bibitem{meltdown}
M.~Lipp, M.~Schwarz, D.~Gruss, T.~Prescher, W.~Haas, A.~Fogh, J.~Horn, S.~Mangard, P.~Kocher, D.~Genkin, Y.~Yarom, and M.~Hamburg, ``Meltdown: reading kernel memory from user space,'' in \emph{Proceedings of the 27th USENIX Conference on Security Symposium}, ser. SEC'18.\hskip 1em plus 0.5em minus 0.4em\relax USA: USENIX Association, 2018, p. 973–990.

\bibitem{OS}
\BIBentryALTinterwordspacing
M.~Mushtaq, M.~M. Yousaf, M.~K. Bhatti, V.~Lapotre, and G.~Gogniat, ``The kingsguard os-level mitigation against cache side-channel attacks using runtime detection,'' \emph{Annals of Telecommunications}, vol.~77, no. 11--12, pp. 731--747, 2022. [Online]. Available: \url{https://doi.org/10.1007/s12243-021-00906-3}
\BIBentrySTDinterwordspacing

\bibitem{MAM_MARIA1}
\BIBentryALTinterwordspacing
M.~Mushtaq, A.~Akram, M.~K. Bhatti, M.~Chaudhry, V.~Lapotre, and G.~Gogniat, ``Nights-watch: a cache-based side-channel intrusion detector using hardware performance counters,'' in \emph{Proceedings of the 7th International Workshop on Hardware and Architectural Support for Security and Privacy}, ser. HASP '18.\hskip 1em plus 0.5em minus 0.4em\relax New York, NY, USA: Association for Computing Machinery, 2018. [Online]. Available: \url{https://doi.org/10.1145/3214292.3214293}
\BIBentrySTDinterwordspacing

\bibitem{MAM_MARIA2}
M.~Mushtaq, J.~Bricq, M.~K. Bhatti, A.~Akram, V.~Lapotre, G.~Gogniat, and P.~Benoit, ``Whisper: A tool for run-time detection of side-channel attacks,'' \emph{IEEE Access}, vol.~8, pp. 83\,871--83\,900, 2020.

\bibitem{Carna}
\BIBentryALTinterwordspacing
S.~Carn\`{a}, S.~Ferracci, F.~Quaglia, and A.~Pellegrini, ``Fight hardware with hardware: Systemwide detection and mitigation of side-channel attacks using performance counters,'' \emph{Digital Threats}, vol.~4, no.~1, Mar. 2023. [Online]. Available: \url{https://doi.org/10.1145/3519601}
\BIBentrySTDinterwordspacing

\bibitem{ALAM}
\BIBentryALTinterwordspacing
M.~Alam, S.~Bhattacharya, and D.~Mukhopadhyay, ``Victims can be saviors: A machine learning--based detection for micro-architectural side-channel attacks,'' \emph{J. Emerg. Technol. Comput. Syst.}, vol.~17, no.~2, Jan. 2021. [Online]. Available: \url{https://doi.org/10.1145/3439189}
\BIBentrySTDinterwordspacing

\bibitem{poly}
N.~F. Polychronou, P.-H. Thevenon, M.~Puys, and V.~Beroulle, ``Madman: Detection of software attacks targeting hardware vulnerabilities,'' in \emph{2021 24th Euromicro Conference on Digital System Design (DSD)}, 2021, pp. 355--362.

\bibitem{Choudhari}
\BIBentryALTinterwordspacing
A.~Choudhari, S.~Guilley, and K.~Karray, ``Specdefender: Transient execution attack defender using performance counters,'' in \emph{Proceedings of the 2022 Workshop on Attacks and Solutions in Hardware Security}, ser. ASHES'22.\hskip 1em plus 0.5em minus 0.4em\relax New York, NY, USA: Association for Computing Machinery, 2022, p. 15–24. [Online]. Available: \url{https://doi.org/10.1145/3560834.3563830}
\BIBentrySTDinterwordspacing

\bibitem{hpc_limitations}
S.~Das, J.~Werner, M.~Antonakakis, M.~Polychronakis, and F.~Monrose, ``Sok: The challenges, pitfalls, and perils of using hardware performance counters for security,'' in \emph{2019 IEEE Symposium on Security and Privacy (SP)}, 2019, pp. 20--38.

\bibitem{challenges}
\BIBentryALTinterwordspacing
M.~Mushtaq, P.~Benoit, and U.~Farooq, ``{Challenges of Using Performance Counters in Security Against Side-Channel Leakage},'' in \emph{{CYBER 2020 - 5th International Conference on Cyber-Technologies and Cyber-Systems}}, Nice, France, Oct. 2020. [Online]. Available: \url{https://hal.science/hal-02979362}
\BIBentrySTDinterwordspacing

\bibitem{yuval}
\BIBentryALTinterwordspacing
W.~Kosasih, Y.~Feng, C.~Chuengsatiansup, Y.~Yarom, and Z.~Zhu, ``Sok: Can we really detect cache side-channel attacks by monitoring performance counters?'' in \emph{Proceedings of the 19th ACM Asia Conference on Computer and Communications Security}, ser. ASIA CCS '24.\hskip 1em plus 0.5em minus 0.4em\relax New York, NY, USA: Association for Computing Machinery, 2024, p. 172–185. [Online]. Available: \url{https://doi.org/10.1145/3634737.3637649}
\BIBentrySTDinterwordspacing

\bibitem{RISCV}
L.~Gerlach, D.~Weber, R.~Zhang, and M.~Schwarz, ``A security risc: Microarchitectural attacks on hardware risc-v cpus,'' in \emph{2023 IEEE Symposium on Security and Privacy (SP)}, 2023, pp. 2321--2338.

\bibitem{mah1}
M.~Khan, M.~Mushtaq, R.~Pacalet, and L.~Apvrille, ``Side-channel attack detection using gem5 and machine learning: A case study on fault-based attacks in risc-v,'' in \emph{2025 IEEE 31st International Symposium on On-Line Testing and Robust System Design (IOLTS)}, 2025, pp. 1--5.

\bibitem{mah3}
------, ``Assessing security risc: Analyzing flush+ fault attack on risc-v using gem5 simulator,'' in \emph{International Conference on Security and Cryptography (SECRYPT)}, 2025.

\bibitem{palumbo}
A.~Palumbo, ``Machine learning-based detection of microarchitectural attacks on risc-v via gem5,'' in \emph{Proceedings of TechDefense 2025 - IEEE International Workshop on Technologies for Defense and Security}, Rome, Italy, November 2025, pp. 1--6, {https://hal.science}{hal-05288651}.

\bibitem{hassan}
\BIBentryALTinterwordspacing
M.~Hassan, M.~Mushtaq, J.~Raik, and T.~Ghasempouri, ``Drsam: Detection of fault-based microarchitectural side-channel attacks in risc-v using statistical preprocessing and association rule mining,'' 2025. [Online]. Available: \url{https://arxiv.org/abs/2510.18612}
\BIBentrySTDinterwordspacing

\bibitem{RISCV_HPCs}
A.-T. Le, T.-T. Hoang, B.-A. Dao, A.~Tsukamoto, K.~Suzaki, and C.-K. Pham, ``Spectre attack detection with neutral network on risc-v processor,'' in \emph{2022 IEEE International Symposium on Circuits and Systems (ISCAS)}, 2022, pp. 2467--2471.

\bibitem{MAIN_ATTACK}
M.~Khan, M.~E. Bin Mohd~Shahfie, M.~Mushtaq, R.~Pacalet, and L.~Apvrille, ``Microarchitectural espionage: Fpga-based security analysis of branch prediction in risc-v out-of-order cores,'' in \emph{2026 14th International Symposium on Digital Forensics and Security (ISDFS)}, 2026, pp. 1--7.

\bibitem{rule1}
T.~Ghasempouri, S.~Payandeh~Azad, B.~Niazmand, and J.~Raik, ``An automatic approach to evaluate assertions' quality based on data-mining metrics,'' in \emph{2018 IEEE International Test Conference in Asia (ITC-Asia)}, 2018, pp. 61--66.

\bibitem{reza}
M.~R.~H. Iman, P.~Chikul, G.~Jervan, H.~Bahsi, and T.~Ghasempouri, ``Anomalous file system activity detection through temporal association rule mining.'' in \emph{ICISSP}, 2023, pp. 733--740.

\bibitem{support}
T.~Ghasempouri, S.~Payandeh~Azad, B.~Niazmand, and J.~Raik, ``An automatic approach to evaluate assertions' quality based on data-mining metrics,'' in \emph{2018 IEEE International Test Conference in Asia (ITC-Asia)}, 2018, pp. 61--66.

\bibitem{verilator}
W.~Snyder, ``{Verilator: The Fastest Verilog/SystemVerilog Simulator},'' \url{https://www.veripool.org/verilator}, 2023, accessed: 2024.

\bibitem{boom}
J.~Zhao, B.~Korpan, A.~Gonzalez, and K.~Asanovic, ``Sonic{BOOM}: The 3rd generation {Berkeley} out-of-order machine,'' in \emph{Fourth Workshop on Computer Architecture Research with RISC-V (CARRV)}, May 2020.

\bibitem{mah_paper}
M.~Khan, M.~E. Bin Mohd~Shahfie, M.~Mushtaq, R.~Pacalet, and L.~Apvrille, ``Microarchitectural espionage: Fpga-based security analysis of branch prediction in risc-v out-of-order cores,'' in \emph{2026 14th International Symposium on Digital Forensics and Security (ISDFS)}, 2026, pp. 1--7.

\bibitem{MAIP}
\BIBentryALTinterwordspacing
M.~Awais, M.~Mushtaq, L.~Naviner, F.~Bruguier, P.~Benoit, and J.~Haj-Yahya, ``Leveraging gem5 and machine learning for end-to-end detection of cache-based side-channel attack patterns,'' in \emph{PROOFS 2025 - 13th International Workshop on Security Proofs for Embedded Systems}, Kuala Lumpur, Malaysia, September 2025, hAL ID: hal-05372979. [Online]. Available: \url{https://hal.science/hal-05372979}
\BIBentrySTDinterwordspacing

\bibitem{MKIP}
M.~Khan, M.~Mushtaq, R.~Pacalet, and L.~Apvrille, ``Side-channel attack detection using gem5 and machine learning: A case study on fault-based attacks in risc-v,'' in \emph{2025 IEEE 31st International Symposium on On-Line Testing and Robust System Design (IOLTS)}, 2025, pp. 1--5.

\bibitem{Le_et_al}
T.-T. Hoang, B.-A. Dao, A.~P. Tsukamoto, K.~Suzaki, and C.-K. Pham, ``Spectre attack detection with neutral network on risc-v processor,'' in \emph{Proceedings of the 2022 IEEE International Symposium on Circuits and Systems (ISCAS)}, Austin, TX, USA, 2022, pp. 2467--2471.

\end{thebibliography}




\end{document}